\documentclass[a4paper,12pt]{extarticle}

\usepackage[english]{babel}
\usepackage[utf8]{inputenc}
\usepackage{amsmath}
\usepackage{amsfonts}
\usepackage{amsthm}
\usepackage{graphicx}
\usepackage{array}
\usepackage[top=2.5cm,bottom=2.5cm,left=2.5cm,right=2.5cm]{geometry}

\newcommand{\C}[3]{C^{#1}\left(#2\middle|#3\right)}
\newcommand{\Col}[4]{C_\mathrm{#1}^{#2}\left(#3\middle|#4\right)}
\newcommand{\qC}[3]{\hat{\mathfrak{C}}^{#1}\left(#2\middle| #3\right)}
\newcommand{\J}[3][]{J_{#1}\left(#2\middle|#3\right)}
\newcommand{\basis}[4]{\theta^{#1}_{#2}\left(#3\middle|#4\right)}
\newcommand{\D}{\mathrm{d}}
\newcommand{\B}{\mathfrak{B}}
\newcommand{\oB}{\widetilde{\mathfrak{B}}}
\newcommand{\cq}[3]{\mathrm{q}_{#1}^{\left(#2,#3\right)}}
\newcommand{\T}{\mathcal{T}_M}
\newcommand{\opc}{\hat{\mathfrak{c}}}

\newcommand{\oG}{\widetilde{\Gamma}}
\newcommand{\oN}{\widetilde{N}}
\newcommand{\oeta}{\widetilde{\eta}}
\newcommand{\oO}{\widetilde{\Omega}}
\newcommand{\Q}{\mathrm{Q}}
\newcommand{\oQ}{\widetilde{\mathrm{Q}}}
\newcommand{\onu}{\widetilde{\nu}}

\begin{document}

\begin{flushright}
\vspace{1mm}
 FIAN/TD/2016-25\\
\end{flushright}

\vskip1.5cm

 \begin{center}
 {\large\bf
 Higher-spin fields and charges\\in the periodic spinor space}
 \vglue 0.6  true cm

\vskip0.5cm

Y.O.~Goncharov and M.A.~Vasiliev

 \vglue 0.3  true cm

 I.E.Tamm Department of Theoretical Physics, Lebedev Physical
 Institute,\\
 Leninsky prospect 53, 119991, Moscow, Russia

 \end{center}

\begin{center}
\vspace{0.5cm}
goncharov@lpi.ru, vasiliev@lpi.ru \\

\par\end{center}

\numberwithin{equation}{section}

\begin{abstract}
\noindent
The $sp\left(2M\right)$ invariant unfolded system  is considered in the
periodic twistor-like spinor space. Complete set of non-trivial charges corresponding to the
global symmetry compatible with the periodicity conditions
is constructed. Residual infinite-dimensional symmetry is
 realized in terms of the star-product algebra. It is shown that
 charges associated with integrations over different cycles are related by
 particular higher-spin symmetry transformations.
\end{abstract}

\section{Introduction}
The  $sp\left(2M\right)$ invariance of the higher-spin (HS) field
multiplet was first proposed in \cite{Fronsdal:1985pd}. The idea that HS theories should
 admit a description in a larger manifestly $sp(2M)$ invariant space-time is as
natural as the idea to describe supersymmetric theories in superspace.
Formulations of HS theories in $sp\left(2M\right)$ invariant (super)spaces
 has been widely elaborated
 (see \cite{Bandos:1999qf,Bandos:1999pq,Vasiliev:2001zy,Vasiliev:2001dc,Vasiliev:2002fs,Bandos:2002te,Didenko:2003aa,PluSorTsu,spspace,BanPasSorTon,BanBekAzSorTsu,IvLuk,Iv,theta,sph,twistors,Sorokin} and references therein).
 In this setup  free massless HS bosonic and fermionic
 fields are described \cite{Vasiliev:2001zy,Vasiliev:2001dc} by a scalar field $C\left(X\right)$ and a svector field
 $C_A\left(X\right)$, respectively, in the generalized space-time $\mathcal{M}_M$
 with local coordinates $X^{AB}=X^{BA}$, $A,B = 1\ldots M$.

   Conserved charges corresponding to conformal and higher symmetries were constructed in
   \cite{Vasiliev:2002fs} (see also \cite{theta,rank,GSV,ads}).
   Unfolded dynamics approach to the $sp(2M)$ invariant equations
    was first considered in
   \cite{Vasiliev:2001zy} and later extended in \cite{theta,sph,twistors} to conserved
   currents and charges. As usual in unfolded dynamics, to this end the
    generalized space-time $\mathcal{M}_M$
    is extended by auxiliary twistor-like spinor variables $Y^A$ to
    $\mathcal{M}_M\times\mathbb{R}^M$. The variables $Y^A$
    together with derivatives
    $\frac{\partial}{\partial Y^A}$  form Heisenberg algebra $H_M$ \cite{sph}
\begin{equation}
\left[\dfrac{\partial}{\partial Y^A},Y^B\right] = \delta_A^B\,
\end{equation}
while the bilinears of oscillators
\begin{equation}
P_{AB} = \dfrac{\partial}{\partial Y^A}\dfrac{\partial}{\partial Y^B},\qquad
K^{AB} = Y^A Y^B, \qquad L_A{}^B = \dfrac{1}{2}\left(Y^A\dfrac{\partial}{\partial Y^B} + \dfrac{\partial}{\partial Y^B} Y^A\right)
\end{equation}
 form $sp\left(2M\right)$. Here $P_{AB}$ and $K^{AB}$ represent generalized
 translations and special conformal transformations. The $gl(M)$ subalgebra
 spanned by $L_A{}^B$ decomposes into generalized dilatation generator $D = L_A{}^A$ and
 $sl(M)$ representing generalized Lorentz transformations generated by
 $l_A{}^B = L_A{}^B - \frac{1}{M}\delta_A^B\, D$ \cite{Vasiliev:2001zy,Vasiliev:2002fs}.

 In this paper we construct a complete set of conserved charges
 in the case of periodic coordinates $Y^A$. Thus,
 the full space where fields live is $\mathcal{M}_M\times T^M$. Analogous problem
 for the non-compact twistor-space
 was considered in \cite{twistors}.

Complete set of non-trivial conserved charges is constructed together with the residual global symmetries they correspond to.
Conserved charges are represented as integrals
 of  closed forms independent of local variations of the integration cycle.
 Despite considerable similarity with the non-compact case, periodicity in
 the twistor-like variables causes a number of peculiarities. One is that, since
 toric geometry allows inequivalent cycles
 non-contractible to each other,  corresponding integrations
 give different sets of conserved charges.
 An interesting output of this paper is that nevertheless
 the latter are related to each other by some HS transformation.
 The complete set of charges can be obtained starting from some elementary cycle
 in the spinor  space.
 Charges associated with other cycles result from those in the spinor space
 by virtue of  higher symmetries. In fact, this implies that HS symmetries can affect
 topology of cycles.

The global symmetry compatible with the periodicity in $Y$ is represented by an
 infinite-dimensional Lie algebra generated by
  basis elements  $\mathrm{T}^r_{\left(\xi,n\right)}$ with $\xi^A\in\left[0,2\pi\right)$, $n_B\in\mathbb{Z}$ and $r=0,1$ obeying the following commutation relations
\begin{equation}\label{eq:residual_Lie}
\left[\mathrm{T}^q_{\left(m,\xi\right)},\mathrm{T}^r_{\left(n,\zeta\right)}\right] = \mathrm{T}^{\left|q+r\right|_2}_{\left(\left(-\right)^r m+n,\left(-\right)^r \xi+\zeta\right)}e^{i\left(-\right)^r \left(m_C\zeta^C - n_C\xi^C\right)} - \mathrm{T}^{\left|q+r\right|_2}_{\left(m+\left(-\right)^q n,\xi+\left(-\right)^q\zeta\right)}e^{-i\left(-\right)^q \left(m_C\zeta^C - n_C\xi^C\right)},
\end{equation}
where $\left|q+r\right|_2:=\left(q + r\right)\mod 2$. Subalgebra of \eqref{eq:residual_Lie} with $q=r=0$ obeys commutation relations
\begin{equation}\label{eq:residual_Lie_subalgebra}
\left[\mathrm{T}^0_{\left(m,\xi\right)},\mathrm{T}^0_{\left(n,\zeta\right)}\right] = 2i\,\sin\left(m_C\zeta^C - n_C\xi^C\right)\, \mathrm{T}^0_{\left(m+n,\xi+\zeta\right)}
\end{equation}
and is somewhat  analogous to the sine algebra introduced in \cite{Fairley} which
is reproduced at $\xi,\zeta\in\mathbb{Z}^M$. Analogously to \cite{Fairley}, relations \eqref{eq:residual_Lie} admit oscillator representation
\begin{equation}
\mathrm{T}^r_{\left(n,\xi\right)}\left(k;v\right) = K^r\star e^{i\xi\, k + in\,v},
\end{equation}
$k\in\mathbb{Z}^M$ and $v^C\in\left[0,2\pi\right)$, with respect to the Moyal-like star product
\begin{multline}
\left(f \star g\right)\left(k;v\right) = \dfrac{1}{\left(2\pi\right)^{2M}}\sum_{m,n\in\mathbb{Z}^M}\int_0^{2\pi}\D^Mu\,\D^Mw \cdot\\\cdot f\left(k+m;v+u\right)g\left(k+n;v+w\right)\exp\left[i\left(m_C w^C - n_C u^C \right)\right],
\end{multline}
acting on functions $f\left(k;v\right) = \sum_N f_N\left(k\right)\,e^{iN_C v^C}$
with half of arguments discrete and another half periodic.
Klein operator $K$ (see e.g. \cite{NonLinHSmanual}) is defined to fulfill the following properties
\begin{equation}
K\star K = 1,\quad K\star f\left(k;v\right) = f\left(-k;-v\right)\star K.
\end{equation}
Note that
$\left[k_C,e^{iN_B v^B}\right]_\star = -2N_C\,e^{iN_B v^B}$, where
$\left[f,g\right]_\star \equiv f\star g - g\star f$.

Riemann theta-function \cite{Mumford}
\begin{equation}
\Theta\left(Y\middle| X\right) = \sum_{n\in\mathbb{Z}^M} \,\exp\left[i\pi\,n_A X^{AB} n_B +2\pi i\,n_A Y^A\right]
\end{equation}
 can be interpreted as an evolution operator (or $\mathcal{D}$-function) for fields, propagating in the $X,Y$ space with periodic $Y$-variables \cite{theta}.
  Connection with field theory may lead to an alternative interpretation of some of
  the important theta-function identities \cite{Mumford,KZ} (see also \cite{theta}) and, other way around, to applications of the apparatus of toric geometry to  field theory.

Consideration of the periodic twistor-like space can open a way toward a uniform description
 of black holes in various dimensions. Black-hole solutions   like Schwarzshild,
 Kerr-Newman, Reisner-Nordström solutions in asymptotically flat
or $AdS$  spacetime or like BTZ black hole in $3d$
are exact solutions of theories of gravity. Moreover BTZ black hole is locally $AdS_3$
resulting from quotiening of the $AdS_3$ group $\mathrm{O}\left(2,2\right)$ over
some discrete subgroup \cite{BTZ}.
Hence it is described by an $AdS_3$ flat connection obeying certain
 periodic boundary conditions. Such construction gives a hint that black holes in various dimensions may be constructed with the aid of BTZ-like solutions in $\mathcal{M}_M\times T^M$ for properly chosen flat connections in unfolded equations. Periodicity of solutions in generalized space-time $\mathcal{M}_M$ is induced by periodicity in spinor variables via unfolded dynamics and black holes themselves perhaps could be obtained as projection of aforementioned ``flat'' solutions onto surfaces in $\mathcal{M}_M$ representing usual space-time. This conjecture provides the main motivation for the problem addressed in this paper, opening a vast area for further research.

The rest of the paper is organized as follows.  In Section $2$
the main ingredients such as fields and currents, their unfolded equations of motion
and symmetry transformations are introduced following
 the non-compact case \cite{twistors}. In Section $3$
 the periodicity conditions  on twistor-like variables are imposed.
 In Section $4$ construction of conserved charges is described and on-shell current
 cohomology along with the full set of non-zero conserved charges are presented.
 In section $5$ it is shown that charges resulting from integration over
  non-homotopic cycles turn out to be related by the action of HS symmetry.
  In Section $6$ conserved charges are represented as symmetry generators acting on
  quantized fields and the full symmetry of dynamics in the periodic spinor space is
   formulated in terms of an infinite-dimensional Lie algebra. In Conclusion
   some peculiar   features of symmetries in the periodic spinor space are discussed.

\section{Fields and currents}

\subsection{Fields}

As shown in \cite{Vasiliev:2001dc} (infinite towers of)
conformal fields in various dimensions ($d\geq 4$) can be
conveniently described in terms of generalized space-time
$\mathcal{M}_M$ with symmetric real matrix coordinates $X^{AB} = X^{BA}$
 ($A,B = 1...M$). For the unfolded formulation $\mathcal{M}_M$
 is extended by auxiliary twistor-like variables $Y^A$ spanning $\mathbb{R}^M$ \cite{Vasiliev:2001zy}.
 Conformal fields are described by scalar functions $C^\pm\left(Y\middle| X\right)$
 obeying rank-one unfolded equations \cite{twistors,rank}

\begin{equation}\label{eq:unfolded_r1}
\left(\dfrac{\partial}{\partial X^{AB}} \pm i\dfrac{\partial ^2}{\partial Y^A \partial Y^B}\right)C^\pm\left(Y\middle| X\right) = 0.
\end{equation}
Equation \eqref{eq:unfolded_r1} expresses covariant constancy condition with the flat connection
\begin{equation}
W^\pm\left(Y,\partial_Y\middle| X\right) = \pm i\,\D X^{AB}\dfrac{\partial^2}{\partial Y^A \partial Y^B}.
\end{equation}
Bosons are described by even functions ($\C{\pm}{-Y}{X} = \C{\pm}{Y}{X}$) while fermions are described by odd ones ($\C{\pm}{-Y}{X} = -\C{\pm}{Y}{X}$) \cite{twistors}.

Unfolded formulation is useful in many respects. In particular, equation \eqref{eq:unfolded_r1} reconstructs $X$-dependence from a given function $\C{\pm}{Y}{0}$,
\begin{equation}
\C{\pm}{Y}{X} = \exp\left[\mp i X^{AB}\dfrac{\partial^2}{\partial Y^A \partial Y^B}\right] \C{\pm}{Y}{0}.
\end{equation}
Fourier decomposition of $\C{\pm}{Y}{0}$ gives the following
 representation for general solution of \eqref{eq:unfolded_r1}
\begin{equation}\label{eq:solution_r1}
\C{\pm}{Y}{X} = \int \D^M\xi\,c^{\pm}\left(\xi\right)\,\exp\left[\pm i\left(\xi_A \xi_B X^{AB} + \xi_B Y^B\right)\right]
\end{equation}
with the elementary solutions
\begin{equation}\label{eq:basis_r1}
\basis{\pm}{\xi}{Y}{X} = \exp\left[\pm i\left(\xi_A \xi_B X^{AB} + \xi_B Y^B\right)\right].
\end{equation}

\noindent As explained in \cite{theta,twistors} (see also \cite{Vasiliev:2001dc}), the superscript $\pm$ in
\eqref{eq:solution_r1} distinguishes between positive- and negative-frequency modes
corresponding to particles and antiparticles upon quantization.
The two modes are complex conjugated
\begin{equation}
\overline{\C{+}{Y}{X}} = \C{-}{Y}{X}\quad \Longleftrightarrow \quad \overline{c^+\left(\xi\right)} = c^-\left(\xi\right).
\end{equation}

Another useful feature of unfolded formulation is that it allows one to describe symmetries of a system in a regular way. Namely, consider a transformation
\begin{equation}
\C{\pm}{Y}{X}\rightarrow \eta\left(Y,\partial_Y\middle| X\right) \C{\pm}{Y}{X}.
\end{equation}
 To be a symmetry, $\eta\left(Y,\partial_Y\middle| X\right)$
should commute with the differential operator on the \textit{lhs} of \eqref{eq:unfolded_r1},
\begin{equation}\label{eq:unfolded_symmetry_r1}
\left[\dfrac{\partial}{\partial X^{AB}} \pm i\dfrac{\partial^2}{\partial Y^A \partial Y^B}, \eta\left(Y,\partial_Y\middle| X\right)\right] = 0.
\end{equation}
Condition \eqref{eq:unfolded_symmetry_r1} is formally consistent since connection in \eqref{eq:unfolded_r1} is flat. The first-order differential operators
\begin{equation}\label{eq:covar_oscillators}
\mathcal{A}_{\pm}{}^C\left(Y\middle|X\right) = Y^C \mp 2i X^{CB}\dfrac{\partial}{\partial Y^B}, \quad \mathcal{B}^{\pm}{}_{C}\left(Y\middle| X\right) = \dfrac{\partial}{\partial Y^C}
\end{equation}
verify \eqref{eq:unfolded_symmetry_r1}. Each pair $\mathcal{A}_+{}^C,\mathcal{B}^+{}_{C}$ and $\mathcal{A}_-{}^C,\mathcal{B}^-{}_{C}$ obeys Heisenberg algebra $H_M$ \cite{sph}
\begin{equation}\label{eq:Heisenberg}
\def\arraystretch{1.5}
\begin{array}{c}
\left[\mathcal{B}^\pm{}_{A},\mathcal{A}_{\pm}{}^B\right] = \delta^B_A\,,\qquad
\left[\mathcal{A}_\pm{}^B,\mathcal{A}_\pm{}^C\right]=0,\quad \left[\mathcal{B}^\pm{}_{B},\mathcal{B}^\pm{}_{C}\right] = 0\,.
\end{array}
\end{equation}
 Since operators \eqref{eq:covar_oscillators} are covariantly constant, they will be referred to as \textit{covariant oscillators}\footnote{Here and after notations of covariant oscillators correspond to \cite{twistors}.}. Any function of covariant oscillators $\eta\left(\mathcal{A}_\pm;\mathcal{B}^\pm\right)$ is a solution of \eqref{eq:unfolded_symmetry_r1} and hence is a symmetry of \eqref{eq:unfolded_r1}.

Covariant oscillators act on \eqref{eq:basis_r1} as follows
\begin{equation}\label{eq:covar_oscillators_action}
\def\arraystretch{1.7}
\begin{array}{l}
\mathcal{B}^\pm{}_{C} \,\theta^\pm_\xi = \pm i\xi_C \,\theta^\pm_\xi,\qquad
\mathcal{A}_{\pm}^C \,\theta^\pm_\xi = \mp i\dfrac{\partial}{\partial \xi_C} \,\theta^\pm_\xi.
\end{array}
\end{equation}
Exponentiation of \eqref{eq:covar_oscillators_action} gives
\begin{equation}\label{eq:covar_oscillators_action_exp}
\def\arraystretch{1.7}
\begin{array}{l}
\exp\left[\pm i\zeta_C\,\mathcal{A}_{\pm}{}^C\right] \,\theta^\pm_\xi = \theta^\pm_{\xi + \zeta},
\end{array}
\end{equation}
allowing to generate the whole basis \eqref{eq:basis_r1} from a single vacuum vector $\theta_0:=\basis{\pm}{0}{Y}{X} = 1$,
\begin{equation}
\def\arraystretch{1.5}
\begin{array}{l}
\mathcal{B}^\pm{}_{C} \,\theta_0 = 0,\qquad
\exp\left[\pm i\xi_C\,\mathcal{A}_{\pm}^C\right] \,\theta_0 = \theta^\pm_{\xi}.
\end{array}
\end{equation}
Any solution to \eqref{eq:unfolded_r1} can  thus be written as
\begin{equation}
\C{\pm}{Y}{X} = \int \D^M\xi\,c^{\pm}\left(\xi\right)\,\exp\left[\pm i\xi_C\,\mathcal{A}_{\pm}^C\right] \,\theta_0.
\end{equation}
 The result of the action of a symmetry transformation can be represented as
\begin{equation}
\eta\left(\mathcal{A}_\pm;\mathcal{B}^\pm\right)\,\C{\pm}{Y}{X} = \int \D^M\xi\,c^\pm\left(\xi\right)\, \eta\left(\mp i\partial_{\xi};\pm i\xi\right)\,\basis{\pm}{\xi}{Y}{X}\,.
\end{equation}

Evolution of a particular field configuration in $X$-variables from $\C{\pm}{Y}{X^\prime}$ to $\C{\pm}{Y}{X}$ is given by a $\mathcal{D}$-function
via the following transformation \cite{twistors}
\begin{equation}
\C{\pm}{Y}{X} = \int \D^M Y^\prime\,\mathcal{D}^\pm\left(Y-Y^\prime\middle| X - X^\prime\right)\,\C{\pm}{Y^\prime}{X^\prime}\,.
\end{equation}
The $\mathcal{D}$-function is a solution to \eqref{eq:unfolded_r1} with the
$\delta$-functional initial data
\begin{equation}
\left(\dfrac{\partial}{\partial X^{AB}} \pm i\dfrac{\partial^2}{\partial Y^A \partial Y^B}\right)\mathcal{D}^\pm\left(Y-Y^\prime\middle| X - X^\prime\right) = 0,\quad \mathcal{D}^\pm\left(Y-Y^\prime\middle| 0\right) = \delta\left(Y - Y^\prime\right).
\end{equation}
Hence,
\begin{equation}
\mathcal{D}^\pm\left(Y\middle| X\right) = \dfrac{1}{\left(2\pi\right)^M}\int \D^M\xi\,\basis{\pm}{\xi}{Y}{X}.
\end{equation}

\subsection{Currents}

Doubling of spinor variables leads to a \textit{rank-two unfolded equation}  \cite{rank}.
\begin{equation}\label{eq:unfolded_r2}
\left(\dfrac{\partial}{\partial X^{AB}} + i \dfrac{\partial^2}{\partial Y_1^A \partial Y_1^B} - i\dfrac{\partial^2}{\partial Y_2^A \partial Y_2^B}\right) \J{Y_1,Y_2}{X} = 0.
\end{equation}
Its solutions $\J{Y_1,Y_2}{X}$ are called \textit{current fields} or simply \textit{currents}. They describe conserved currents
 in the unfolded formulation. The flat connection
\begin{multline}
W^{(2)}\left(Y_{1,2},\partial_{Y_{1,2}}\middle| X \right) = i\,\D X^{AB}\left( \dfrac{\partial^2}{\partial Y_1^A \partial Y_1^B} - \dfrac{\partial^2}{\partial Y_2^A \partial Y_2^B}\right) = W^+\left(Y_{1},\partial_{Y_{1}}\middle| X \right) + W^-\left(Y_{2},\partial_{Y_{2}}\middle| X \right)
\end{multline}
is the sum of flat connections for positive- and negative-frequency modes of
 \eqref{eq:unfolded_r1}.
Hence bilinears of rank-one fields
\begin{equation}\label{eq:bilinear}
\J{Y_{1,2}}{X} = \C{+}{Y_1}{X}\C{-}{Y_2}{X}\,,
\end{equation}
called \textit{bilinear currents},
verify \eqref{eq:unfolded_r2}. The straightforward generalization of currents \eqref{eq:bilinear}
 via extension of the set of rank-one fields \eqref{eq:solution_r1} by a color index $\mathrm{i}=1...\mathcal{N}$
\begin{equation}\label{eq:bilinear_color}
\J{Y_{1,2}}{X} = \sum_{\mathrm{i}=1}^{\mathcal{N}}\Col{i}{+}{Y_1}{X}\Col{i}{-}{Y_2}{X},
\end{equation}
plays  the central role in $AdS\slash CFT$ correspondence
(see \textit{e.g.} \cite{KlebPol,Giombi:2012ms}) and were considered for the non-compact
twistor-like space \textit{e.g.} in  \cite{twistors}. Since it does not play any role in  this
paper and can be easily reinserted at any moment we will not consider it in the sequel.

Bilinear current \eqref{eq:bilinear} is a particular case of a
more general current field
\begin{equation}
\label{eq:bilinear_general}
\J[\eta]{Y_{1,2}}{X} = \eta\left(Y_{1,2},\partial_{Y_{1,2}}\middle| X\right)\C{+}{Y_1}{X}\C{-}{Y_2}{X},
\end{equation}
where $\eta\left(Y_{1,2},\partial_{Y_{1,2}}\middle| X\right)$ is a symmetry of \eqref{eq:unfolded_r2}. Analogously to the rank-one case $\eta$ commutes with covariant differential of \eqref{eq:unfolded_r2}
\begin{equation}\label{eq:unfolded_symmetry_r2}
\left[\dfrac{\partial}{\partial X^{AB}} + i\dfrac{\partial^2}{\partial Y_1^A \partial Y_1^B} -i\dfrac{\partial^2}{\partial Y_2^A \partial Y_2^B}, \eta\left(Y_{1,2},\partial_{Y_{1,2}}\middle| X\right)\right] = 0.
\end{equation}
 Covariant oscillators \eqref{eq:covar_oscillators}
\begin{equation}\label{eq:covar_oscillators_r2}
\mathcal{A}_+{}^C\left(Y_1\middle| X\right),\mathcal{B}^+{}_{C}\left(Y_1\middle| X\right)\quad \text{and}\quad \mathcal{A}_-{}^C\left(Y_2\middle| X\right),\mathcal{B}^-{}_{C}\left(Y_2\middle| X\right)
\end{equation}
verify \eqref{eq:unfolded_symmetry_r2}, hence any function
$\eta\left(\mathcal{A}_{+},\mathcal{A}_{-};\mathcal{B}^{+},\mathcal{B}^{-}\right)$ is a symmetry of
\eqref{eq:unfolded_r2} and the most general form of a bilinear current is \cite{twistors}
\begin{equation}
\label{eq:bilinear_general_oscillators}
\J[\eta]{Y_{1,2}}{X} = \eta\left(\mathcal{A}_{+},\mathcal{A}_{-};\mathcal{B}^{+},\mathcal{B}^{-}\right)\C{+}{Y_1}{X}\C{-}{Y_2}{X}.
\end{equation}

The action of covariant oscillators on the rank-two basis vectors $\basis{+}{\xi}{Y_1}{X}\basis{-}{\zeta}{Y_2}{X}$
\begin{equation}
\def\arraystretch{1.5}
\begin{array}{ll}
\mathcal{B}^+{}_{C}\,\theta^+_\xi\theta^-_\zeta = i\xi_C\,\theta^+_\xi\theta^-_\zeta, & \mathcal{B}^-{}_{C}\,\theta^+_\xi\theta^-_\zeta = -i\zeta_C\,\theta^+_\xi\theta^-_\zeta,\\
\mathcal{A}_+{}^C\,\theta^+_\xi\theta^-_\zeta = -i\dfrac{\partial}{\partial \xi_C}
\theta^+_\xi\theta^-_\zeta, & \mathcal{A}_-{}^C\,\theta^+_\xi\theta^-_\zeta =
 i\dfrac{\partial}{\partial \zeta_C} \theta^+_\xi\theta^-_\zeta
\end{array}
\end{equation}
 generates the complete basis from a single vacuum vector $\theta^{(2)}_0 := \theta^+_0 \theta^-_0 = 1$,
\begin{equation}\label{eq:basis_covar_r2}
\def\arraystretch{1.5}
\begin{array}{l}
\mathcal{B}^\pm{}_{C}\,\theta^{(2)}_0 = 0\,,\qquad
\theta^+_\xi \theta^-_\zeta=\exp\left[i\xi_C\,\mathcal{A}_+{}^C - i\zeta_C\,\mathcal{A}_-{}^C\right]\,\theta^{(2)}_0 .
\end{array}
\end{equation}
The action of a symmetry parameter in \eqref{eq:bilinear_general_oscillators} is
\begin{equation}
\label{eq:bilinear_general_oscillators_explicit}
\J[\eta]{Y_{1,2}}{X} = \int \D^M\xi\D^M\zeta\,c^+\left(\xi\right)c^-\left(\zeta\right)\,\eta\left(-i\partial_{\xi},i\partial_{\zeta};i\xi,-i\zeta\right)\basis{+}{\xi}{Y_1}{X}\basis{-}{\zeta}{Y_2}{X}.
\end{equation}

It is convenient to introduce the following linear combinations of
the covariant oscillators $\mathcal{A},\mathcal{B}$
\begin{equation}\label{eq:covar_essential}
\begin{array}{ll}
\B_C = \mathcal{B}^-{}_{C} -\mathcal{B}^+{}_{C}, & \oB_C = \mathcal{B}^-{}_{C} +\mathcal{B}^+{}_{C},\\
\oB^C = \dfrac{1}{2}\left(\mathcal{A}_-{}^C - \mathcal{A}_+{}^C\right), &
\B^C = \dfrac{1}{2}\left(\mathcal{A}_-{}^C + \mathcal{A}_+{}^C\right)\\
\end{array}
\end{equation}
with the  non-zero commutation relations
\begin{equation}
\big[ \B_A,\oB^B\big] = \delta_A^B,\quad\big[\oB_A,\B^B\big] = \delta_A^B.
\end{equation}
These oscillators are most conveniently represented as differential operators
\begin{equation}
\def\arraystretch{1.5}
\begin{array}{ll}
\B_C = \dfrac{\partial}{\partial U^C}, & \oB_C = \dfrac{\partial}{\partial V^C},\\
\oB^C = U^C + iX^{CB}\dfrac{\partial}{\partial V^B}, &
\B^C = V^C + iX^{CB}\dfrac{\partial}{\partial U^B}
\end{array}
\end{equation}
in terms of the variables \cite{twistors}
\begin{equation}\label{eq:uv}
Y_1 = V - U,\quad Y_2 = V + U.
\end{equation}

Let us introduce an involutive antiautomorphism $\rho$ of the algebra of covariant oscillators that acts as follows
\begin{equation}\label{eq:antiautomorphism}
\rho\left(\mathcal{A}_\pm{}^B\right) =\mathcal{A}_{\mp}{}^B,\quad \rho\left(\mathcal{B}^\pm{}_C\right) = -\mathcal{B}^\mp{}_C.
\end{equation}
The oscillatros $\B,\oB$ are $\rho$-even and $\rho$-odd, respectively,
\begin{equation}
\rho\left(\mathfrak{B}\right) =\mathfrak{B},\quad \rho (\oB) = -\oB.
\end{equation}

For practical computations it is convenient to chose a specific ordering
prescription for functions of covariant oscillators. We will use the
totally symmetric Weyl ordering described by the Weyl star product.
In these terms, any two symbols ({\it i.e.,} functions of commuting variables)
$f\left(\mathcal{A}_+,\mathcal{A}_-;\mathcal{B}^+,\mathcal{B}^-\right)$ and
$g\left(\mathcal{A}_+,\mathcal{A}_-;\mathcal{B}^+,\mathcal{B}^-\right)$
are star-multiplied as follows
\begin{equation}\label{eq:starproduct}
\left(f * g\right)\left(\mathcal{A};\mathcal{B}\right) =
f\left(\mathcal{A};\mathcal{B}\right)\,\exp{\dfrac{1}{2}\sum_{a=+,-}
\left(\dfrac{\overleftarrow{\partial}}{\partial \mathcal{B}^a{}_{C}}\dfrac{\overrightarrow{\partial}}{\partial \mathcal{A}_a{}^C} - \dfrac{\overleftarrow{\partial}}{\partial \mathcal{A}_a{}^{C}}\dfrac{\overrightarrow{\partial}}{\partial \mathcal{B}^a{}_{C}}\right)}\; g\left(\mathcal{A};\mathcal{B}\right).
\end{equation}
In terms of the star product (\ref{eq:starproduct})
the vacuum vector $\theta^{(2)}_0$ obeying $\mathcal{B}^\pm{}_{C}*\theta_0^{(2)} = 0$
is realized as
\begin{equation}
\theta^{(2)}_0 = \exp\left[-2\sum_{a=+,-}\mathcal{A}_a{}^C \mathcal{B}^a{}_{C} \right].
\end{equation}
Symbols of the basis star-product elements $\theta^+_\xi\theta^-_\zeta$ can be
generated from the vacuum $\theta^{(2)}_0$
by the left star-multiplication  \eqref{eq:starproduct} via
\eqref{eq:basis_covar_r2}
\begin{multline}
\theta^+_\xi \theta^-_\zeta = \exp\left[i\xi_C\,\mathcal{A}_+{}^C - i\zeta_C\,\mathcal{A}_-{}^C\right]*\theta_0^{(2)} = \exp\left[2i\xi_C\,\mathcal{A}_+{}^C - 2i\zeta_C\,\mathcal{A}_-{}^C\right]\cdot\theta_0^{(2)}.
\end{multline}
In terms of the star product, symmetry parameters are represented by their symbols
$\eta\left(\mathcal{A};\mathcal{B}\right)$. Since Weyl ordering is totally symmetric,
antiautomorhism $\rho$ \eqref{eq:antiautomorphism} acts on a symbol
$f\left(\mathcal{A}_+,\mathcal{A}_-;\mathcal{B}^+,\mathcal{B}^-\right)$ simply as
\begin{equation}
\rho\left(f\left(\mathcal{A}_+,\mathcal{A}_-;\mathcal{B}^+,\mathcal{B}^-\right)\right) = f\left(\mathcal{A}_-,\mathcal{A}_+;-\mathcal{B}^-,-\mathcal{B}^+\right).
\end{equation}
Indeed, it is straightforward to check that $\rho$ is  an antiautomorphism of the star-product
algebra \eqref{eq:starproduct}, {\it i.e.} $\rho\left(f*g\right) = \rho\left(g\right) * \rho\left(f\right)$.

\section{Periodic spinor space}

To construct periodic solutions it suffices to put \eqref{eq:solution_r1}  on a lattice by setting $\xi_A = \dfrac{2\pi}{\ell^{(A)}}n_A$ with $n_A\in\mathbb{Z}$ (or in condensed notation $\xi = \dfrac{2\pi}{\ell}n$ for $n\in\mathbb{Z}^M$)
\begin{equation}\label{eq:solution_r1_periodic}
\C{\pm}{Y}{X} = \dfrac{\left(2\pi\right)^M}{\ell^{(1)}...\ell^{(M)}}\;\sum_n\, c\left(\dfrac{2\pi}{\ell}n\right)\,\basis{\pm}{2\pi\,n\slash\ell}{Y}{X}\,,
\end{equation}
\begin{equation}\label{eq:basis_r1_periodic_raw}
\basis{\pm}{2\pi\,n\slash\ell}{Y}{X} = \exp\left[\pm i\left(4\pi ^2\,n_A n_B\,\dfrac{X^{AB}}{\ell^{(A)}\ell^{(B)}} + 2\pi\,n_A\,\dfrac{Y^A}{\ell^{(A)}} \right)\right].
\end{equation}
Such solutions are $\ell^{(A)}$-periodic in $Y^A$-variables.
The non-compact limit corresponds to  $\ell^{(A)}\to\infty$.

It is convenient
to use rescaled variables and change notations as follows
\begin{equation}\label{eq:rescaled}
Y^{\prime A} := \dfrac{2\pi}{\ell^{(A)}}\,Y^A,\; X^{\prime AB} :=
\dfrac{4\pi^2}{\ell^{(A)}\ell^{(B)}}\,X^{AB}\,.
\end{equation}
Since this is equivalent to setting
\begin{equation}\label{lpi}
\ell^{(A)}=2\pi,
\end{equation}
 in the sequel we will
not distinguish between primed and unprimed variables. The dependence
on $\ell^{(A)}$ can be easily reconstructed in the very end if necessary.

In terms of rescaled variables  basis vectors \eqref{eq:basis_r1_periodic_raw}
 are
\begin{equation}\label{eq:basis_r1_periodic}
\basis{\pm}{n}{Y}{X} := \exp\left[\pm i\left(n_A n_B\, X^{AB} + n_B\, Y^B\right)\right]\,,
\end{equation}
\noindent
while any periodic solution \eqref{eq:solution_r1_periodic} can be written as follows
\begin{equation}\label{eq:solution_r1_periodic_norm}
\C{\pm}{Y}{X} = \sum_{n\in\mathbb{Z}^M} c^\pm_n\,\basis{\pm}{n}{Y}{X}.
\end{equation}
Basis functions \eqref{eq:basis_r1_periodic} (and hence functions \eqref{eq:solution_r1_periodic_norm}) are $2\pi$-periodic in
$Y$-variables,  $2\pi$-periodic in $X^{AA}$ and $\pi$-periodic in $X^{AB}$ with $A\neq B$. Hence, unfolded dynamics induces periodicity in $\mathcal{M}_M$
from that in the spinor variables. It also implies that, reintroducing arbitrary radii,
periods of the $X$-variables factorize into products of periods of $Y$-variables.
Namely, periods of  $Y$- and $X$-variables are $\ell^{(A)}$ for $Y^A$,
$\frac{\ell^{(A)}\ell^{(A)}}{2\pi}$ for $X^{AA}$ and
$\frac{\ell^{(A)}\ell^{(B)}}{4\pi}$ for $X^{AB}$ ($A\neq B$).
So,  possible
 periods of the $\frac{M\left(M+1\right)}{2}$-dimensional space $\mathcal{M}_M$,
 that can be respected by solutions of the rank-one equations \eqref{eq:unfolded_r1},
 are parametrized by $M$ numbers.

Due to the second relation in \eqref{eq:covar_oscillators_action}
which does not respect periodicity,
  polynomials of covariant oscillators $\mathcal{A}_\pm{}^C$ (\ref{eq:covar_oscillators})
  do not act properly on \eqref{eq:basis_r1_periodic}.
 The  generators respecting  periodicity are
\begin{equation}\label{eq:symmetries_Fock_r1_periodic}
\def\arraystretch{1.5}
\begin{array}{l}
\mathcal{B}^\pm{}_{C} \theta^\pm_n = \pm i\, n_C \,\theta^\pm_n,\qquad
\exp\left[\pm i\, m_C\,\mathcal{A}_\pm{}^C\right] \theta^\pm_n = \theta^\pm_{n+m}
\end{array}
\end{equation}
for any $m,n\in \mathbb{Z}^M$.
As in the non-compact case, basis vectors are generated from a single vacuum vector $\theta_0$
\begin{equation}
\def\arraystretch{1.5}
\begin{array}{l}
\mathcal{B}^\pm{}_{C}\, \theta_0 = 0\,,\qquad
\exp\left[\pm i\, n_C\,\mathcal{A}_\pm{}^C\right] \theta_0 = \theta^\pm_{n}.
\end{array}
\end{equation}
 Periodicity  demands any symmetry transformation to be $2\pi$-periodic in the
  oscillators $\mathcal{A}$
\begin{equation}
\eta = \eta\left(e^{i\,\mathcal{A}_\pm};\mathcal{B}^\pm\right).
\end{equation}
The parameter $\eta $ can be viewed as a polynomial of $\mathcal{B}^\pm$ and  a Laurent polynomial of $e^{i\,\mathcal{A}_\pm}$. The action of a symmetry transformation can be written as
\begin{equation}
\eta\left(e^{i\,\mathcal{A}_\pm};\mathcal{B}^\pm\right)\,\C{\pm}{Y}{X} = \sum_n\, c^\pm_n\,\eta\left(e^{\pm\frac{\partial}{\partial n}};in\right)\,\basis{\pm}{n}{Y}{X}.
\end{equation}

For rank-two equation \eqref{eq:unfolded_r2} periodic Ansatz is introduced in the same
manner. Bilinear currents \eqref{eq:bilinear} are built from positive- and negative-frequency rank-one fields \eqref{eq:solution_r1_periodic_norm}
\begin{equation}
\J{Y_{1,2}}{X} = \sum_{m,n} c^+_m c^-_n\, \basis{+}{m}{Y_1}{X}\basis{-}{n}{Y_2}{X}.
\end{equation}
Basis elements $\basis{+}{m}{Y_1}{X}\basis{-}{n}{Y_2}{X}$ are constructed from the
 vacuum vector $\theta^{(2)}_0$ analogously to \eqref{eq:basis_covar_r2}
\begin{equation}\label{eq:basis_covar_r2_periodic}
\def\arraystretch{1.5}
\begin{array}{l}
\mathcal{B}^\pm{}_{C}\,\theta^{(2)}_0 = 0\,,\qquad
\exp\left[i m_C\,\mathcal{A}_+{}^C - i n_C\,\mathcal{A}_-{}^C\right]\,\theta^{(2)}_0 = \theta^+_m \theta^-_n.
\end{array}
\end{equation}
In terms of $Y_1,Y_2$ and $U,V$ \eqref{eq:uv} they have the form
\begin{multline}\label{eq:basis_r2_periodic}
\basis{+}{m}{Y_1}{X}\basis{-}{n}{Y_2}{X} = \exp\left[i\left(\left(m+n\right)_B\left(m-n\right)_C \,X^{BC} + m_C\, Y_1^C - n_C\, Y_2^C \right)\right] =\\= \exp\left[i\left(\left(m+n\right)_B\left(m-n\right)_C \,X^{BC} - \left(m+n\right)_C\, U^C + \left(m-n\right)_C\, V^C \right)\right].
\end{multline}
 Periodicity properties  of  the vector \eqref{eq:basis_r2_periodic} in $X^{AB}$
 are the same  as of \eqref{eq:basis_r1_periodic}.

The global symmetry transformation respects periodicity in $Y_1$ and $Y_2$ variables
 iff they are generated by  $e^{i\,\mathcal{A}_a{}^C}$ and $\mathcal{B}^a{}_{C}$,
\begin{equation}\label{eq:symmetry_r2_periodic}
\eta = \eta\left(e^{\pm i\,\mathcal{A}_{a}};\mathcal{B}^b\right)\,,\qquad a,b = +,-\,.
\end{equation}
Within the Weyl ordering the periodic star-product symbols of parameters \eqref{eq:symmetry_r2_periodic} admit Fourier decomposition,
\begin{equation}\label{eq:Fourier_decompose}
\eta\left(\mathcal{A};\mathcal{B}\right) =  \sum_{k,l\in\mathbb{Z}^M}\eta_{kl}\left(\mathcal{B}^+,\mathcal{B}^-\right)
\,e^{ik_B\,\mathcal{A}_+{}^B} e^{il_C\,\mathcal{A}_-{}^C}.
\end{equation}
This gives the following explicit formula for the symmetry action
\begin{multline}\label{eq:symmetry_action_periodic}
\J[\eta]{Y}{X} := \eta\left(\mathcal{A};\mathcal{B}\right)\,*\,\J{Y}{X} = \sum_{m,n,k,l} c_m^+ c_n^-\,\eta_{kl}\left(im + \dfrac{ik}{2},-in+\dfrac{il}{2}\right)\cdot\theta_{m+k}^+ \theta_{n-l}^-.
\end{multline}
In terms of oscillators \eqref{eq:covar_essential} decomposition \eqref{eq:Fourier_decompose} is
\begin{equation}\label{eq:Fourier_decompose_essential_raw}
\eta\big(\B,\oB\big) =
\sum_{k,l} \eta_{kl}\big(\B_C,\oB_D\big)\, e^{i\left(k+l\right)_A\,\B^A}
 e^{-i\left(k-l\right)_B\,\oB^B}.
\end{equation}
Let, for $N_A\in\mathbb{Z}$,  $|N_A|_2 = N_A\mod 2$, and for $N\in\mathbb{Z}^M$, $\left|N\right|_2\in\mathbb{Z}_2{}^M$ is understood component-wise. Then $|k+l|_2 = |k-l|_2$  and hence
 decomposition \eqref{eq:Fourier_decompose_essential_raw}  can be rewritten as follows
\begin{equation}\label{eq:Fourier_decompose_essential}
\eta\big(\B,\oB\big) = \sum_{|N|_2 = |\oN|_2} \eta_{N,\oN}\big(\B_C,\oB_D\big)\, e^{iN_A\,\B^A} e^{i\oN_B\,\oB^B}.
\end{equation}

$\mathcal{D}$-functions for periodic solutions of \eqref{eq:unfolded_r1} can be introduced
analogously to the non-compact case \cite{theta}. In the positive-frequency sector,
the $\mathcal{D}$-function
\begin{equation}\label{eq:theta_raw}
\theta\left(Y\vert X\right) = \dfrac{1}{\left(2\pi\right)^M}\sum_n\,\exp\left[i\left(n_A n_B\,X^{AB} + n_A\,Y^A\right)\right]
\end{equation}
is  a solution to \eqref{eq:unfolded_r1} with the $\delta$-functional initial data on a torus,
\begin{equation}
\theta\left(Y\vert 0\right) = \delta\left(Y\right),\quad Y^A\in \left[-\pi,\pi\right)\;\text{for $A = 1...M$}.
\end{equation}
\noindent Along with $Y$-periodicity (and aforementioned $X$-periodicity) it is
 quasi-periodic  with $X$ being the matrix of quasi-periods \cite{Mumford}
\begin{equation}
\theta\left(Y^C + 2m_B \,X^{BC}\middle|X\right) = e^{-i\left(m_B m_C\, X^{BC} + m_B\,Y^B\right)}\,\theta\left(Y\middle|X\right).
\end{equation}
Up to simple  redefinitions of arguments
expression \eqref{eq:theta_raw}  represents Riemann theta-function \cite{Mumford}
\begin{equation}
\Theta\left(Y\middle| X\right) = \sum_{n\in\mathbb{Z}^M}\exp\left[i\pi\, n_A n_B\,X^{AB} + 2\pi i\,n_A Y^A\right].
\end{equation}
Action of covariant oscillators $e^{ib\, \mathcal{B}} e^{ia\,\mathcal{A}}\,
\theta\left(Y\middle|X\right)$ gives rise to theta-functions with rational
characteristics $a_C\in\mathbb{Q}$ and $b^C\in\mathbb{Q}$ ($C=1...M$) (\cite{Mumford},
 see also \cite{KZ})
\begin{equation}
\Theta_{a,b}\left(Y\middle|X\right) := \sum_n\,\exp\left[i\pi \left(n + a\right)_B\left(n + a\right)_C X^{BC} + 2\pi i\,\left(n + a\right)_B\, \left(Y + b\right)^B\right].
\end{equation}

\section{Charges}
\subsection{Charge components and integration surfaces}

Conserved charges can be represented as integrals
of on-shell-closed current differential forms.
Current forms are constructed from  an arbitrary rank-two field \cite{theta,twistors,ads},
and, in particular, from the  bilinear currents (\ref{eq:bilinear_general_oscillators}).
In the non-compact case  the closed on-shell current $M$-form is \cite{twistors} (see also \cite{theta})
\begin{equation}\label{eq:charge_form}
\Omega\left(J_\eta\right) = W^1\wedge ... \wedge  W^M\,\left.\J[\eta]{Y_{1,2}\left(U,V\right)}{X}\right|_{U=0},
\end{equation}
with $U,V$   \eqref{eq:uv}. $ W^A$ is the operator-valued 1-form
\begin{equation}\label{eq:w}
 W^A = \D V^A + i\,\D X^{AB}\,\dfrac{\partial}{\partial U^B}.
\end{equation}
Conserved charges result from integration over an $M$-dimensional surface
$\Sigma \subset \mathcal{M}_M\times \mathbb{R}^M$ which is  spacelike in
$X$-variables \cite{twistors}
($\mathbb{R}^M$ is parametrized by spinor variables $V^A$),
\begin{equation}\label{eq:charge}
\mathrm{Q}_\eta = \int_\Sigma \Omega\left(J_\eta\right).
\end{equation}
Charge \eqref{eq:charge} is independent of  local variations of $\Sigma$ since
 $\D\Omega\left(J_\eta\right) = 0$ by virtue of the current equation.
 Non-trivial charges correspond to the on-shell de Rham cohomology of the set of forms \eqref{eq:charge_form}. As presented in \cite{twistors}, in the non-compact case non-zero charges are completely represented by the  $\oB$-independent
  symmetry parameters $\eta\left(\mathfrak{B}\right)$. In other words, given
  $\eta\big(\mathfrak{B},\oB\big)$ there exists such $\eta^\prime\left(\mathfrak{B}\right)$ that $\Omega\left(J_{\eta^\prime}\right) - \Omega\left(J_\eta\right) = \D\omega$.

Another set of dual closed current forms $\oO\left(J_{\oeta}\right)$
 is constructed via exchange $U\leftrightarrow V$ \cite{twistors}.
  Nontrivial conserved charges for such current forms are represented
   by $\oeta\big(\oB\big)$. Hence the complete set of charges is doubled giving rise to
 the  $\mathcal{N} = 2$ supersymmetric HS algebra \cite{twistors}.
 For definiteness in this paper we mostly focus on current forms \eqref{eq:charge_form}.
 The dual set of charges can be considered analogously.

For the $Y$-periodic case the situation is somewhat different.
Now functions \eqref{eq:solution_r1_periodic_norm}  live on a torus
$\T\times T^M := \left(\mathcal{M}_M\times \mathbb{R}^M\right)\slash L$ where
$L\subset \mathcal{M}_M\times \mathbb{R}^M$ is the lattice corresponding to the periods of rescaled coordinates \eqref{eq:rescaled}
\begin{equation}\label{eq:lattice}
L = \left\{Y^A = 2\pi\,p,\, X^{AA} = 2\pi\,q,\, \left. X^{AB}\right|_{A\neq B} = \pi\,r\;\text{for}\; p,q,r\in \mathbb{Z}\right\}\,.
\end{equation}
Integration surfaces $\Sigma$, being $M$-dimensional cycles in $\T\times T^M$, may
  belong to different homotopy classes. These are anticipated to generate different charges.

In the periodic case, the question whether it is possible to
eliminate the $\oB$-dependence from the
 symmetry parameters \eqref{eq:Fourier_decompose_essential_raw}
 has to be reconsidered. The goal is to find the essential part of $\B$- and $\oB$-dependence of the symmetry
parameters, associated with the current cohomology in the periodic case.

For symmetry parameters \eqref{eq:Fourier_decompose_essential_raw} and  periodic
solutions \eqref{eq:solution_r1_periodic_norm} the current forms
 \eqref{eq:charge_form} are
\begin{multline}\label{eq:charge_form_periodic}
\Omega\left( J_\eta\right) = \sum_{m,n,k,l}\left(\D\,V^A + \left(m+n+k-l\right)_C \D\,X^{CA}\right)^{\wedge M} \\
 c_m^+ c_n^-\,\eta_{kl}\left(-i\left(m+n+\frac{k-l}{2}\right), i\left(m-n+\frac{k+l}{2}\right)\right)\cdot\\ \cdot \exp\left[i\left(m-n+k+l\right)_B\left(V^B + \left(m+n+k-l\right)_D \,X^{DB}\right)\right]\,,
\end{multline}
where $\left( W^A\right)^{\wedge M} :=  W^{1}\wedge ...\wedge  W^{M}$. Current form \eqref{eq:charge_form_periodic} is defined on
$\mathcal{T}_M\times T^M$ where the twistor-like sector $T^M$ is parametrized by variables $V^A\in\left[0,2\pi\right)$.
 Integration of \eqref{eq:charge_form_periodic} over a compact surface $\Sigma\subset\mathcal{T}_M\times T^M$ gives
\begin{equation}\label{eq:form_integration}
\int_\Sigma \Omega\left( J_\eta\right) =
\sum_{m,n,k,l} c_m^+ c_n^-\,\eta_{kl}\left(-i\left(m+n+\frac{k-l}{2}\right), i\left(m-n+\frac{k+l}{2}\right)\right)\cdot \cq{\Sigma}{m-n+k+l}{m+n+k-l},
\end{equation}
where
\begin{equation}\label{eq:charge_components}
\cq{\Sigma}{\nu}{\onu} := \int_{\Sigma} \mathrm{d}w^1\wedge ... \wedge \mathrm{d}w^M
\,e^{i\nu_C\,w^C}\,,\qquad w^A := V^A + \onu_C \,X^{CA}
\end{equation}
will be referred to as \textit{charge components}
  and
\begin{equation}\label{eq:nunu}
\nu_C = \left(m-n+k+l\right)_C,\quad \onu_C = \left(m+n+k-l\right)_C.
\end{equation}
Charge components are independent of local variations of $\Sigma$ since the differential form in \eqref{eq:charge_components} is  closed.

Integration in \eqref{eq:charge} and \eqref{eq:charge_components} should be
 performed over \textit{space-like $M$-cycles}. Space-like directions in $\mathcal{M}_M$ are associated with the traceless parts of $X^{AB}$ \cite{Vasiliev:2001dc}
\begin{equation}\label{eq:space_time}
\sum_{A=1}^M X^{AA} = 0.
\end{equation}
Consider the following parametrization of $\mathcal{M}_M$ by variables $t,y_{i,j+1},z_i$
($i\le j$ and $i,j=1...M-1$):
\begin{equation}\label{eq:parametrization_M}
\begin{array}{c}
X^{11} = z_1,\,X^{22} = -z_1 + z_2,...,X^{M-1,M-1} = - z_{M-2}+z_{M-1},X^{MM} = t - z_{M-1},\\
 \quad y_{i,j+1} = X^{i,j+1} = X^{j+1,i}.
\end{array}
\end{equation}
 Here $t = \sum_{A=1}^M X^{AA}$ parametrizes time while $y$ and $z$, parametrizing
the traceless part of $X^{AB}$, are space coordinates.
Note that transformation \eqref{eq:parametrization_M} is from  $\mathrm{SL}\left(\frac{M(M+1)}{2}\middle|\mathbb{Z}\right)$. Therefore it preserves the lattice \eqref{eq:lattice} acting properly on the torus $\T\subset  \mathcal{M}_M$.

The freedom  in the choice of parametrization of space-like directions in \eqref{eq:space_time}, \textit{i.e.} of traceless part of $X^{AB}$ is not essential. Different parametrizations resulting from $\mathrm{SL}\left(\frac{M(M+1)}{2}\middle|\mathbb{Z}\right)$ transformations of $X^{AB}$ preserve
the lattice and  give equivalent sets of conserved charges. Indeed, in this case  fundamental cycles corresponding to one parametrization are expressed as integer combinations of those for the other. The same is true for conserved charges being integrals over $M$-dimensional space-like cycles in $\T\times T^M$. More generally, parametrizations of $\T\times T^M$ resulting from $\mathrm{SL}\left(\frac{M\left(M+1\right)}{2} + M\middle| \mathbb{Z}\right)$ coordinate transformations of $X^{AB},Y^A$  give equivalent charges.

For instance, consider parametrization \eqref{eq:parametrization_M}.
Consider fundamental space-like $M$-cycles $\left\{\sigma_\mathsf{a}\right\}$ of $\T\times T^M$ with a single winding
parametrized by all sets of $M$ pairwise different $y$, $z$ and $V$.
A single-winding cycle $\sigma_0$ in the spinor space parametrized by the variables $V$
 will be referred to as the \textit{lower cycle}.
Other space-like cycles will be called \textit{higher}.
Any space-like cycle $\Sigma$ is homotopic to a sum of fundamental cycles
\begin{equation}\label{eq:fundamental_cycles}
\Sigma = \sum_\mathsf{a} b_\mathsf{a} \sigma_\mathsf{a}
\end{equation}
with the coefficients $b_\mathsf{a}\in\mathbb{Z}$  representing the number of windings
over the respective fundamental cycle.

Charge components \eqref{eq:charge_components} being
linear functions on the space of cycles are determined by their values
$\cq{\sigma_\mathsf{a}}{\nu}{\onu}$ on the fundamental cycles. Using \eqref{eq:parametrization_M} for
\eqref{eq:charge_components} and setting $t=0$ one arrives at the sum of monomials of the $M$-th power in $\D V,\D y,\D z$  which correspond to integration over space-like fundamental cycles. One can see that for any $\sigma_\mathsf{a}$
\begin{equation}\label{eq:charge_components_calculated}
\cq{\sigma_\mathsf{a}}{\nu}{\onu} \propto p_{\sigma_\mathsf{a}}\left(\onu\right)\,\delta_{\nu,0}
\,,\qquad \delta_{\nu,0} := \delta_{\nu_1,0}\,...\,\delta_{\nu_M,0}\,,
\end{equation}
where  $p_{\sigma_\mathsf{a}}\left(\onu\right)$ is a homogeneous polynomial of $\onu_A$.
Indeed, formula \eqref{eq:charge_components_calculated} results from the change of variables $w^A$ in \eqref{eq:charge_components} to those among $V,y,z$ \eqref{eq:parametrization_M} that parametrize $\sigma_\mathsf{a}$, with $p_{\sigma_\mathsf{a}}\left(\onu\right)$ being the Jacobian. As a result,
for any cycle
$\sigma_\mathsf{a}$ there is a linear transformation
$F_\mathsf{a}\left[\onu\right]$ of variables $\nu$ leading to
\eqref{eq:charge_components_calculated} in the form
\begin{equation}\label{eq:cycles_det}
\cq{\sigma_\mathsf{a}}{\nu}{\onu} \propto \det F_\mathsf{a}\left[\onu\right] \,\delta_{F_\mathsf{a}\left[\onu\right]\nu,0}.
\end{equation}
Generally, different cycles may give the same polynomials.
Because charge components \eqref{eq:charge_components} are linear functions of cycles,
for any cycle $\Sigma$ (\ref{eq:fundamental_cycles})
\begin{equation}\label{eq:charge_components_cycles}
\cq{\Sigma}{\nu}{\onu} \propto p_{\Sigma}\left(\onu\right)\,\delta_{\nu,0}
\,,\qquad
p_\Sigma = \sum_{\mathsf{a}} b_\mathsf{a} p_{\sigma_\mathsf{a}}\,.
\end{equation}
A useful consequence of \eqref{eq:charge_components_cycles} is
the expression for
charge components of any cycle in terms of those for the lower fundamental one
\begin{equation}\label{eq:to_V}
\cq{\Sigma}{\nu}{\onu} \propto p_{\Sigma}\left(\onu\right)\,\cq{\sigma_0}{\nu}{\onu}\,.
\end{equation}
Note that $p_{\sigma_0}\left(\onu\right) \propto 1$.

\subsection{ $M=2$ example }

\noindent As an example, consider the $M=2$ case in some detail. General parametrization \eqref{eq:parametrization_M} for $X^{AB}$ is
\begin{equation}\label{eq:parametrization_M2}
X = t\begin{pmatrix}
1 & 0\\
0 & 0\\
\end{pmatrix} + y\begin{pmatrix}
0 & 1\\
1 & 0\\
\end{pmatrix} + z\begin{pmatrix}
1 & 0\\
0 & -1\\
\end{pmatrix}.
\end{equation}
The volume form in \eqref{eq:charge_components} for $t=0$ is
\begin{multline}\label{eq:volume_M2}
\D w^1\wedge \D w^2 =
\left(\D\, V^1 + \onu_2\,\D\, y + \onu_1\,\D \, z\right)\wedge \left(\D\, V^2 + \onu_1\,\D\,y - \onu_2\,\D\,z\right) =\\=
\D\,V^1\wedge \D \,V^2 + \onu_1\,\D\,V^1\wedge \D\,y - \onu_2\,\D\,V^2\wedge \D\,y -\\- \onu_2\,\D\, V^1\wedge \D\,z - \onu_1\, \D\,V^2\wedge \D\,z - \left(\onu_1{}^2 + \onu_2{}^2\right)\,\D\, y\wedge \D \, z.
\end{multline}
With the particular parametrization \eqref{eq:parametrization_M2}
fundamental $2$-cycles of $\T\times T^M$ are associated with the following pairs of variables
$V^1V^2$ (the lower cycle) , $V^1y$ , $V^2z$, $V^2y$, $V^1z$ and $yz$.

Charge components for $\sigma_0 = V^1V^2$  are
\begin{equation}
\cq{\sigma_0}{\nu}{\onu} = \iint_{0}^{2\pi} \D V^1\D V^2\;e^{i\nu_1\,V^1 + i\nu_2\,V^2} = 4\pi^2\,\delta_{\nu_1,0}\delta_{\nu_2,0}\propto \delta_{\nu,0}.
\end{equation}
Analogous computation for  $V^1y$ gives
\begin{equation}
\cq{V^1y}{\nu}{\onu} = \onu_1 \int_0^\pi\D y \int_{0}^{2\pi} \D V^1\;e^{i\nu_1\,V^1 + i\big(\nu_1\onu_2 + \onu_1\nu_2\big)\,y} = 2\pi^2\,\onu_1\,\delta_{\nu_1,0}\delta_{\nu_1\onu_2 + \nu_2\onu_1,0},
\end{equation}
In  agreement with \eqref{eq:charge_components_calculated}
this is equivalent to
\begin{equation}
\mathrm{q}_{V^1y}^{\left(\nu,\onu\right)} \propto\onu_1\,\delta_{\nu,0}\,.
\end{equation}
Analogous computation of the full set of charge components for fundamental cycles gives
\begin{equation}
\begin{array}{c}
\mathrm{q}_{V^1V^2}^{\left(\nu,\onu\right)} \propto \delta_{\nu,0},\\
\mathrm{q}_{V^1y}^{\left(\nu,\onu\right)} \propto\onu_1\,\delta_{\nu,0},\quad \mathrm{q}_{V^2z}^{\left(\nu,\onu\right)} \propto \onu_1\,\delta_{\nu,0},\\
\mathrm{q}_{V^2y}^{\left(\nu,\onu\right)} \propto \onu_2\,\delta_{\nu,0},\quad \mathrm{q}_{V^1z}^{\left(\nu,\onu\right)} \propto \onu_2\,\delta_{\nu,0},\\
\mathrm{q}_{yz}^{\left(\nu,\onu\right)} \propto\left(\onu_1{}^2 +\onu_2{}^2\right)\delta_{\nu,0}.
\end{array}
\end{equation}
The respective polynomials in \eqref{eq:charge_components_cycles} are integer combinations of
 $1$, $\onu_1$, $\onu_2$ and $\onu_1{}^2 + \onu_2{}^2$.

\subsection{On-shell current cohomology and non-zero charges}

\subsubsection{$\oB_C$-dependence}

Here we show that, analogously to the non-compact case \cite{twistors} the dependence
on $\oB_C$ can be eliminated  from the parametrization of non-zero
conserved charges. Namely, for a given symmetry parameter $\eta_{kl}\big(\B_A,\oB_B\big)$ we introduce another parameter
\begin{equation}\label{eq:equivalent_parameter}
\eta^\prime_{kl}\big(\B_A\big) = \eta_{kl}\big(\B_A,-i\frac{k+l}{2}\big)
\end{equation}
depending solely on $\B_A$ such that current forms \eqref{eq:charge_form_periodic}
with $\eta_{kl}\big(\B_A,\oB_B\big)$ and
 $\eta^\prime_{kl}\big(\B_A\big)$ differ by an exact form.

Indeed, using  \eqref{eq:charge_form_periodic}, \eqref{eq:charge_components}, \eqref{eq:nunu}
and \eqref{eq:equivalent_parameter}, 
\begin{multline}\label{eq:forms_difference}
\Omega\left(J_\eta\right) - \Omega\left(J_{\eta^\prime}\right) = \sum_{m,n,k,l}\left(\D w^A\right)^{\wedge M}
 c_m^+ c_n^- \exp\left[i\nu_B w^B\right]\\ \Delta\eta_{kl}\left(-i\left(m+n+\frac{k-l}{2}\right), i\left(m-n+\frac{k+l}{2}\right)\right) \,,
\end{multline}
where 
\begin{multline}
\Delta\eta_{kl}\left(-i\left(m+n+\frac{k-l}{2}\right), i\left(m-n+\frac{k+l}{2}\right)\right)=\\=\eta_{kl}\left(-i\left(m+n+\frac{k-l}{2}\right),i\left(m-n+\frac{k+l}{2}\right)\right) -
\eta^\prime_{kl}\left(-i\left(m+n+\frac{k-l}{2}\right)\right) =\\= \int_0^1 \D t\;\dfrac{\D}{\D t} \eta_{kl}\left(-i\left(m+n+\frac{k-l}{2}\right),-i\frac{k+l}{2} + it\,\nu\right)=\\= i\nu_C\int_0^1 \D t\,\frac{\partial \eta_{kl}}{\partial \oB_C}\left(-i\left(m+n+\frac{k-l}{2}\right),-i\frac{k+l}{2} + it\,\nu\right).
\end{multline}
Hence 
\begin{equation}
\Omega\left(J_\eta\right) - \Omega\left(J_{\eta^\prime}\right) = \D\beta,
\end{equation}
where
\begin{multline}
\beta \propto \sum_{m,n,k,l} c_m^+ c_n^-\,\varepsilon_{A_1...A_M}\,\D w^{A_1}\wedge ... \wedge \D w^{A_{M-1}}\,e^{i\nu_C\, w^C}\cdot\\ \cdot \int_0^1 \D t\,\frac{\partial \eta_{kl}}{\partial \oB_{A_M}}\left(-i\left(m+n+\frac{k-l}{2}\right),-i\frac{k+l}{2} + it\,\nu\right).
\end{multline}

\subsubsection{$\B^C$- and $\oB^C$-dependence}
 Symmetry parameters \eqref{eq:Fourier_decompose_essential} with reduced
 $\oB_A$-dependence are
\begin{equation}\label{eq:parameters}
\eta\big(\B_C;\B^A,\oB^B\big) = \sum_{|N|_2 = |\oN|_2} \eta_{N,\oN}\left(\B_C\right)\, e^{iN_A\,\B^A} e^{i\oN_B\,\oB^B}.
\end{equation}
The non-compact case \cite{twistors} suggests that oscillators $\B^A$
play the central role in parametrization of non-trivial charges.
On the other hand, since for a symmetry parameter
\eqref{eq:Fourier_decompose_essential} periodicity implies that $|N|_2 = |\oN|_2$, the
$\oB^A$-dependence cannot be fully eliminated in the periodic case. Indeed, parameters
\eqref{eq:parameters} depending solely on parameters $\B$,
\begin{equation}\label{eq:parameters_B}
\eta\big(\B\big) = \sum_{N} \eta_{N}\left(\B_C\right)\, e^{2iN_A\,\B^A},
\end{equation}
give rise to charges for the lower cycle $\sigma_0$ (\textit{cf.} \eqref{eq:form_integration}, \eqref{eq:Fourier_decompose_essential_raw}-\eqref{eq:Fourier_decompose_essential} and \eqref{eq:to_V})
\begin{equation}\label{eq:charge_B}
\mathrm{Q}_{\eta} \propto \sum_{m,n,N} c^+_m c^-_n\,\eta_N\left(-i\left(m+n\right)\right)\,\delta_{m-n+2N,0},
\end{equation}
where $c_m^+$ and $c_n^-$ enter $\Q_\eta$ with $\left|m\right|_2 = \left|n\right|_2$. The latter condition implies that parameters of the form \eqref{eq:parameters_B} do not represent
 the full set of conserved charges. Since the form of charge components \eqref{eq:charge_components_cycles}
 gives $\delta_{m-n+2N,0}$ in \eqref{eq:charge_B} for any cycle $\Sigma$ this is true for the general case.
 In the sequel we focus on integration over the lower cycle $\sigma_0$ showing
 in the next section that this allows us to obtain the full set of non-trivial
 conserved charges.

The $\oB^A$-dependence can be however minimized as follows. For a parameter \eqref{eq:parameters} the conserved charge for $\sigma_0$ is
\begin{equation}
\Q_\eta \propto\sum_{m,n,|N|_2=|\oN|_2} c_m^+ c_n^-\, \eta_{N,\oN}\left(-i\left(m+n-\frac{\oN}{2}\right)\right)\,\delta_{m-n+N,0}.
\end{equation}
\noindent
Let $\eta_{N,\oN}$ have definite grading $\Gamma = |N|_2\in\mathbb{Z}_2{}^M$. Since the charge
depends only on the following combinations of parameters
\begin{equation}
\eta_N\left(-ik\right) = \sum_{\oN:|\oN|_2 = \Gamma} \eta_{N,\oN}\left(-ik + i\,\dfrac{\oN}{2}\right),
\end{equation}
it can be represented by any term with $|\oN|_2=\Gamma$. The simplest options are
\begin{equation}\label{eq:parameter_representative}
\eta^{\left(\pm\Gamma\right)} = \sum_{N:|N|_2 =
\Gamma} \eta_N\left(\mathfrak{B}_C\right)\,e^{iN_A\,\B^A}e^{\pm i\Gamma_B\,\oB^B}\,.
\end{equation}
The antiautomorphism $\rho$ acts on \eqref{eq:parameter_representative} as follows
\begin{equation}\label{eq:rho_action}
\rho\left(\eta^{\left(\pm\Gamma\right)}\right) = \eta^{\left(\mp\Gamma\right)}.
\end{equation}
Since  $\rho$ is involutive, parameters $\eta$ can be decomposed into
$\rho$-even and $\rho$-odd parts
\begin{equation}\label{eq:even_odd}
\eta^\pm = \dfrac{1\pm\rho}{2}\,\eta.
\end{equation}
For the case of $\Gamma = 0$, $\eta^- = 0$ .

Note that star product of two parameters \eqref{eq:parameter_representative} is not
necessarily of the form \eqref{eq:parameter_representative} because, generally,
$\Gamma + \Gamma^\prime\notin\mathbb{Z}_2{}^M$. However, parameters of the
form \eqref{eq:parameter_representative} are not demanded to form an algebra and
will only  be  used for calculation of charges which are
\begin{equation}\label{eq:charge_gamma}
\mathrm{Q}_\eta = \sum_{m,n:|m-n|_2 = \Gamma} c_m^+ c_n^-\, \eta^{\left(\pm\Gamma\right)}_{n-m}\left(-i\left(m+n\right)\right)\,,\qquad
\eta^{\left(\pm\Gamma\right)}_N\left(k\right) = \eta_N\left(k \pm \frac{i\Gamma}{2}\right)\,.
\end{equation}

The grading $\Gamma\in\mathbb{Z}_2{}^M$ can be interpreted as distinguishing
 between bosonic and fermionic degrees of freedom.
 This suggests an extension of the initial periodic spinor Ansatz by allowing anti-periodic
 (Neveu-Schwarz) conditions. Detailed consideration of this issue is, however,
  beyond the scope of this paper.

Non-trivial dual  charges resulting from the substitution
$V\leftrightarrow U$ in \eqref{eq:charge_form} are parametrized by
\begin{equation}\label{eq:parameter_representative_cong}
\oeta^{\left(\oG\right)} = \sum_{\oN:|\oN|_2 = \oG} \oeta_{\oN}\big(\oB_C\big)\,e^{i\oN_A\,\oB^A}e^{i\oG_B\,\B^B}
\end{equation}
having the form
\begin{equation}\label{eq:charges_dual}
\oQ_{\oeta} = \sum_{m,n:|m-n|_2 = \oG} c_m^+ c_n^-\,
\oeta^{\big(\oG\big)}_{m+n}\left(i\left(m-n\right)\right)\,,\qquad \oeta^{\big(\oG\big)}_N\left(k\right) =
\oeta_N\left(k + \frac{i\oG}{2}\right).
\end{equation}
Note that at $\oG = 0$ parameters \eqref{eq:parameter_representative_cong} are $\B$-independent giving $\eta^\pm = 0$ whenever $\eta\big(-\oB\big) = \mp\ \eta\big(\oB\big)$.

\subsection{Non-compact limit}

The  $\mathbb{Z}_2{}^M$-grading $\Gamma$ accounting for even and odd components $N_A$ in
\eqref{eq:parameter_representative} degenerates in the non-compact
 limit $\ell\to\infty$. Indeed,  in terms of  oscillators \eqref{eq:covar_essential},
 the rescaled oscillator $\oB_C$  is $\dfrac{2\pi}{\ell^{(C)}}\oB_C$. Hence
\begin{equation}
e^{i \,\dfrac{2\pi}{\ell^{(A)}}\oB^A} \xrightarrow{\ell\to\infty} 1.
\end{equation}
This repoduces the result of \cite{twistors} for  the non-compact case that  non-trivial charges are parametrized solely by the oscillators $\B$.
Fourier components $c_n$ in \eqref{eq:solution_r1_periodic_norm} reproduce their non-compact
analogs $c\left(\xi\right)$ in \eqref{eq:solution_r1} with $\xi = \dfrac{2\pi}{\ell}n$
\begin{equation}
\dfrac{\left(2\pi\right)^M}{\ell^{(1)} ... \ell^{(M)}}\,\sum_n\,... \xrightarrow{\ell\to\infty} \int \D^M\xi\,...\;.
\end{equation}

Independence of non-compact charges of the integration surface  \cite{twistors}
 is also reproduced in the limit $\ell\to\infty$.
Indeed, charge components  \eqref{eq:charge_components} for fundamental cycles where
shown to be of the form \eqref{eq:cycles_det}.
They are different for different integration cycles because $F_{\mathrm{a}}\left[\onu\right]$
 corresponds to a particular fundamental cycle $\sigma_\mathsf{a}$. In the non-compact limit the
 dependence of charge components on integration cycles disappears since
\begin{equation}\label{eq:limit}
\dfrac{\left(2\pi\right)^M}{\ell^{(1)}...\ell^{(M)}}
\det F\left[\onu\right]\,\delta_{F\left[\onu\right]\nu,0}\xrightarrow{\ell\to\infty}\det
 F\left[\onu\right]\cdot\delta\left(F\left[\onu\right]\nu\right) = \delta\left(\nu\right),
\end{equation}
where $\delta\left(\nu\right)$ on the \textit{r.h.s.} of \eqref{eq:limit} is the Dirac delta-function.

\section{Higher-spin symmetry mappings between different cycles}

An interesting outcome of the developed techniques is that
charges resulting from any cycle are equivalent to
those evaluated on the lower cycle  with appropriately modified charge parameters.
Indeed, dropping $\oB_C$-dependence in (\ref{eq:charge_form_periodic})
gives
\begin{multline}\label{eq:B_independent}
\Omega\left(J_\eta\right) = \sum_{m,n,k,l} \left(\D\,V^A + \left(m+n+k-l\right)_C \D\,X^{CA}\right)^{\wedge M}
 c_m^+ c_n^-\,\eta_{kl}\left(-i\left(m+n+\frac{k-l}{2}\right)\right)\\ \cdot \exp\left[i\left(m-n+k+l\right)_B\left(V^B + \left(m+n+k-l\right)_D \,X^{DB}\right)\right].
\end{multline}
The charge resulting from \eqref{eq:B_independent} by
 integration over any cycle $\Sigma$ equals to a charge associated with
 the lower fundamental cycle $\sigma_0 = V^1...V^M$ with appropriately
 modified symmetry parameter. In more detail, let
\begin{equation}\label{eq:parameter_P}
\eta^\prime_{kl}\left(-i\left(m+n+\dfrac{k-l}{2}\right)\right) = p_\Sigma\left(m+n+k-l\right)\eta_{kl}\left(-i\left(m+n+\frac{k-l}{2}\right)\right).
\end{equation}
Integration  of \eqref{eq:charge_form} with $\eta^\prime$ \eqref{eq:parameter_P}
over $\sigma_0$ gives the same charge as for
$\eta_{kl}\left(-i\left(m+n+\frac{k-l}{2}\right)\right)$  integrated over $\Sigma$. Indeed, using \eqref{eq:to_V}, integration over $\Sigma$ with parameter $\eta$ gives a factor of
  $p_\Sigma\left(m+n+k-l\right)$, while on the \textit{rhs} of \eqref{eq:parameter_P} it is
   included into $\eta^\prime$.

In terms of star product \eqref{eq:starproduct} relation \eqref{eq:parameter_P}
takes the form
\begin{equation}\label{eq:switch}
\eta^\prime\left(\B_C;\B^A,\oB^B\right) = p_\Sigma\left(i\,\B_C\right) * \eta\left(\B_C;\B^A,\oB^B\right).
\end{equation}
This implies that a charge for a higher cycle $\Sigma$ corresponding to some symmetry
$\eta$ equals to a charge for the lower cycle corresponding to the higher symmetry $p_{\Sigma} * \eta$.

Let conserved charges be considered  as pairings between cycles and symmetry parameters
\begin{equation}
\left<\sigma_\mathsf{a},\eta\right> = \int_{\sigma_\mathsf{a}} \Omega\left(J_\eta\right).
\end{equation}
Consider a transformation $\Xi_\mathsf{a}$ mapping the lower cycle $\sigma_0$ to a
higher one
$\sigma_\mathsf{a}=\Xi_\mathsf{a}\left(\sigma_0\right) $.
By \eqref{eq:switch}
\begin{equation}\label{eq:conjugate}
\left<\Xi_\mathsf{a}\left(\sigma_0\right),\eta\right> =
\left<\sigma_0,p_{\sigma_\mathsf{a}} * \eta\right>\,.
\end{equation}
As a result, transformation of symmetry parameters \eqref{eq:switch} is conjugate to a
transition from the lower cycle to $\Sigma$ represented by an integer combination
$\Sigma = \sum_{\mathsf{a}}b_{\mathsf{a}}\,\Xi_\mathsf{a}\left(\sigma_0\right)$ (\textit{cf.} \eqref{eq:fundamental_cycles}).
Note that only specific polynomials $p_{\sigma_{\mathsf{a}}}$ described in \eqref{eq:charge_components_calculated} and their integer combinations generate
transformation \eqref{eq:switch} conjugate to  mappings of the lower cycles to
higher ones.
An important outcome is that any conserved charge can be obtained by integration over
the spinor space, \textit{i.e.} over the lower cycle $\sigma_0$, for some symmetry parameter $\eta\big(\B_C;\B^A,\oB^B\big)$.
 This is somewhat analogous to the situation  in the non-compact case, where, for a
 given parameter $\eta$, conserved charge is independent of the integration cycle.
 In the periodic case  transition to the integration over
 the spinor space is always possible with the symmetry parameters transformed according to
 \eqref{eq:switch} and hence involving HS algebra into play.

 Transformation \eqref{eq:conjugate} relates different geometric structures to algebraic
 properties of the symmetry transformations, \textit{i.e.} higher integration cycles correspond to
 higher symmetries which are naturally included into the whole framework. On the other hand, in the
 customary lower-symmetry framework  there is no room for algebraic relations between
 different integration cycles.

An interesting remaining question is to describe inverse transformation, \textit{i.e.} conditions on the
 symmetry parameters $\eta$ allowing to obtain the
same charge from a higher cycle $\Sigma$ with some symmetry parameter $\eta^\prime$
such that
$
\left<\sigma_0,\eta\right> = \left<\Sigma,\eta^\prime\right>.
$
According to \eqref{eq:conjugate} this is possible provided that
$\eta = p_{\Sigma} * \eta^\prime$. Analysis of this issue is
less trivial, demanding a  definition of a proper class
of (may be nonpolynomial) functions $\eta^\prime$. Its
detailed consideration is beyond the scope
 of this paper.

\section{Algebra of charges and symmetries}

\subsection{Charges as symmetry operators}

Conserved charges correspond to symmetries of the rank-one system \eqref{eq:unfolded_r1}
 via Noether's theorem. Constructed in terms of rank-two fields they can be realized as symmetry
 generators acting on rank-one fields. Via quantization of rank-one fields Fourier amplitudes
 $c_n^\pm$ become operators $\opc_n^\pm$ with non-trivial commutation relations  (see \cite{Vasiliev:2001dc}
 for details of the quantization procedure and \cite{twistors} for algebra of charges).
Analogously to \cite{theta}, the non-zero commutation relations in the periodic case are
\begin{equation}
\left[\opc_m^-,\opc_n^+\right] = \,\delta_{mn}.
\end{equation}
With the symmetry parameters $\eta^{\left(\Gamma\right)}$ \eqref{eq:parameter_representative}
 quantized conserved charges \eqref{eq:charge_gamma} become operators
\begin{equation}\label{eq:charge_operator}
\hat{\mathrm{Q}}_{\eta} = \sum_{m,n} \eta^{\left(\Gamma\right)}_{n-m}
\left(-i\left(m+n\right)\right)\,\opc^+_m\opc^-_n\,.
\end{equation}
As in the non-compact case \cite{twistors} they form closed algebra with respect to commutators
\begin{equation}\label{eq:charge_algebra}
\left[\hat{\mathrm{Q}}_{\eta^\prime},\hat{\mathrm{Q}}_{\eta}\right] =
\hat{\mathrm{Q}}_{\left[\eta^\prime,\eta\right]_\star}\,,
\end{equation}
where
$\eta\left(k;v\right) = \sum_N\,\eta_N\left(k\right) e^{iNv}$ ($k_B\in\mathbb{Z},
v^C\in \left[0,2\pi\right)$) are Weyl symbols for the
 Moyal-like star product
\begin{multline}\label{eq:Moyal}
\left(f \star g\right)\left(k;v\right) = \sum_{m,n\in\mathbb{Z}^M}\int_0^{2\pi}
\dfrac{\D^Mu\,\D^Mw }{\left(2\pi\right)^{2M}} f\left(k+m;v+u\right)g\left(k+n;v+w\right)\exp\left[i\left(m_C w^C - n_C u^C \right)\right],
\end{multline}
and the corresponding star commutator is $\left[f,g\right]_\star = f \star g - g \star f$, $\left[k_C,e^{iN_Bv^B}\right]_\star = -2N_C\, e^{iN_Bv^B}$. One-to-one correspondence between symbols of the form
\begin{equation}\label{eq:symbol}
\eta\left(k;v\right) = \sum_N \eta_N\left(-ik\right)e^{iNv}
\end{equation}
and charges \eqref{eq:charge_operator} results from the substitution $k\leftrightarrow m+n$ and $N\leftrightarrow n-m$ for Fourier components $\eta_N\left(-ik\right)$. The charge $\hat{\mathrm{Q}}_{\left[\eta^\prime,\eta\right]_\star}$ in
\eqref{eq:charge_algebra} is associated with the symbol
$\left[\eta^\prime,\eta\right]_\star\left(k;v\right)$.
 The dual set of charges for parameters $\eta$ is constructed from the
  symbols $K\star \eta$, where $K$ is Klein operator (see e.g. \cite{NonLinHSmanual})
  obeying
\begin{equation}
K\star K = 1,\quad K\star f\left(k;v\right) = f\left(-k;-v\right)\star K.
\end{equation}
In terms of star product \eqref{eq:Moyal} it is represented by the delta-function
\begin{equation}
K = \left(2\pi\right)^M\delta_{k,0}\,\delta\left(v\right).
\end{equation}
The whole algebra of symmetries is thus $\mathbb{Z}_2$-graded by $K$ and parametrized by symbols of the form $\varepsilon = \eta + K\star\eta^\prime$ with symmetry operators obeying commutation relations
\begin{equation}\label{eq:charge_algebra_full}
\left[\hat{\mathrm{Q}}_{\varepsilon^\prime},\hat{\mathrm{Q}}_{\varepsilon}\right] = \hat{\mathrm{Q}}_{\left[\varepsilon^\prime,\varepsilon\right]_\star}.
\end{equation}
This is the straightforward generalization of the non-compact construction of
\cite{twistors}.

It is straightforward to see that for parameters $\eta^{\prime\left(\Lambda\right)}$ and $\eta^{\left(\Gamma\right)}$ with gradings $\Lambda$ and $\Gamma$, their product $\eta^{\prime\left(\Lambda\right)}\star\eta^{\left(\Gamma\right)}$ has grading $\Lambda + \Gamma \mod 2$.

The charges act on quantized rank-one fields $\qC{\pm}{Y}{X} =
\sum_n\,\opc^\pm_n\basis{\pm}{n}{Y}{X}$ by  commutator.
For instance, for the symmetry parameters $\B_C$ and
$\eta_N = e^{iN_B\,\B^B}e^{i|N_C|_2\oB^C}$, the charges are
\begin{equation}
\hat{\mathrm{Q}}_{\B_C} = -2i\sum_n n_C\,\opc^+_n\opc^-_n,\qquad
\hat{\mathrm{Q}}_{\eta_N} =  \sum_n \opc^+_n\opc^-_{n+N}
\end{equation}
acting on $\qC{+}{Y}{X}$ as follows
\begin{equation}
\left[\hat{\mathrm{Q}}_{\B_C},\hat{\mathfrak{C}}^+\right] = -2\sum_n in_C\,\opc^+_n\theta^+_n,\quad \left[\hat{\mathrm{Q}}_{\eta_N},\hat{\mathfrak{C}}^+\right] = \sum_n \,\opc^+_n\theta^+_{n+N}.
\end{equation}
This is equivalent to the action \eqref{eq:symmetries_Fock_r1_periodic}
of operators $-2\,\mathcal{B}^+{}_{C}$ and
$e^{iN_C\,\mathcal{A}_+{}^C}$ on $\C{+}{Y}{X}$.
Computation for $\qC{-}{Y}{X}$ and the operators
$-2\,\mathcal{B}^-{}_{C}$ and $-e^{iN_C\,\mathcal{A}_-{}^C}$ is analogous.
This makes the  correspondence between conserved charges and symmetries of
rank-one system \eqref{eq:unfolded_r1} manifest.

\subsection{Symmetry algebra }

Periodicity of $Y$-variables changes symmetries of the rank-one system compared to
 the non-compact case.
 Residual symmetry algebra, that respects periodicity,
 is presented by conserved charge operators \eqref{eq:charge_algebra_full}  as functionals of symmetry parameters
 \eqref{eq:symbol}, acting on
 quantized rank-one fields via commutator. Symmetries of the
 rank-one system are thus generated by symbols of
 symmetry transformations constituted by monomials of the type
 $K^{r}\star k_{C_1}...k_{C_m}\;e^{ in_B\,v^B}$ ($r=0,1$) which can be packed into
 generating functions
\begin{equation}\label{eq:sin_elements}
\mathrm{T}^r_{\left(n,\xi\right)}\left(k;v\right) = K^r\star e^{i\xi\,k + in\,v},\quad \xi^C\in \left[0,2\pi\right),
\;n\in\mathbb{Z}^M,\,r=0,1
\end{equation}
with the star product \eqref{eq:Moyal}.
Polynomials in $k$'s can be obtained via differentiation over $\xi^B$ at $\xi = 0$.
The set of generating functions \eqref{eq:sin_elements} is closed with respect to the star commutator
\begin{equation}\label{eq:sin_algebra}
\left[\mathrm{T}^q_{\left(m,\xi\right)},\mathrm{T}^r_{\left(n,\zeta\right)}\right]_\star = \mathrm{T}^{\left|q+r\right|_2}_{\left(\left(-\right)^r m+n,\left(-\right)^r \xi+\zeta\right)}e^{i\left(-\right)^r \left(m_C\zeta^C - n_C\xi^C\right)} - \mathrm{T}^{\left|q+r\right|_2}_{\left(m+\left(-\right)^q n,\xi+\left(-\right)^q\zeta\right)}e^{-i\left(-\right)^q \left(m_C\zeta^C - n_C\xi^C\right)}.
\end{equation}
The generators with $q=r=0$ form a subalgebra obeying
\begin{equation}\label{eq:sin_algebra_even}
\left[\mathrm{T}^0_{\left(m,\xi\right)},\mathrm{T}^0_{\left(n,\zeta\right)}\right]_\star = 2i\,\sin\left(m_C\zeta^C - n_C\xi^C\right)\, \mathrm{T}^0_{\left(m+n,\xi+\zeta\right)},
\end{equation}
which is analogous to the sine algebra introduced in \cite{Fairley}, where its oscillator representation
  was also presented. The difference is that a half of indices in
   \eqref{eq:sin_elements} are continuous, while for the sine algebra all
  of them are discrete. Relations \eqref{eq:sin_algebra} obey Jacobi identity and
   hence elements \eqref{eq:sin_elements} form a Lie algebra with respect to star
   commutator \eqref{eq:sin_algebra}. This infinite-dimensional Lie algebra represents
   the symmetry of rank-one system \eqref{eq:unfolded_r1} with periodic  variables $Y$.

\section{Conclusion}

Analysis of conserved charges of the HS equations with periodic  twistor-like
coordinates $Y^A$ performed in this paper exhibits several interesting features.
The charges are represented as integrals of closed current forms in the extended
$X^{AB}$, $Y^A$ space. Since periodicity in the $Y$-variables implies periodicity
in $X$ variables, one can consider charges associated with different cycles
in the $X^{AB}$, $Y^A$ space.

Closed current forms may depend on
symmetry parameters $\eta$  parametrizing different charges like momentum, electric charge,
conformal weight as well as their HS generalizations. In the non-compact case
the most general symmetry parameters $\eta$ depend on the four types of oscillators
 $\mathcal{A}_{\pm}$ and $\mathcal{B}^{\pm}$ \eqref{eq:covar_oscillators}.
 In the periodic case, the symmetry parameters depend on $\mathcal{B}^\pm$ and
 $e^{i\mathcal{A}_\pm}$.

Nontrivial charges are represented by the current cohomology, \textit{i.e.} those
closed current forms that are not exact. In the non-compact case the current
cohomology was shown in \cite{twistors} to be represented by the symmetry parameters
that depend solely on the oscillators  $\B$ or $\oB$ \eqref{eq:covar_essential}.
 In the periodic case the situation is slightly different with the current
 cohomology represented by various $\mathbb{Z}_2{}^M$-graded parameters of the form
\begin{equation}
\eta^{\left(\Gamma\right)} = \sum_{N:|N|_2 =
\Gamma} \eta_N\big(\mathfrak{B}_C\big)\,e^{iN_A\,\B^A}e^{ i\Gamma_B\,\oB^B},\quad \Gamma\in\mathbb{Z}_2{}^M
\end{equation}
and
\begin{equation}
\oeta^{\left(\oG\right)} = \sum_{\oN:|\bar{N}|_2 = \oG} \oeta_{\oN}\big(\oB_C\big)\,e^{i\oN_A\,\oB^A}e^{i\oG_B\,\B^B},\quad \oG\in\mathbb{Z}_2{}^M
\end{equation}
for the dual set of charges.

Another peculiarity of the periodic case is that naive expectation that charges
evaluated as integrals over non-equivalent cycles are different is not quite true.
Namely, the complete set of charges can be obtained by integration
over the lower fundamental cycle $\sigma_0 = V^1...V^M$ constituted
solely by spinor variables.
Other  cycles $\Sigma$ for a given symmetry parameter $\eta$ give charges which can be
also obtained  from the lower cycle with appropriate higher symmetry $\eta^\prime$
\begin{equation}\label{eq:hs_shift}
\eta^\prime = p_\Sigma\left(i\,\B_C\right)*\eta .
\end{equation}
 This means that HS symmetries act on different non-contractible to each other cycles and hence connect them algebraically.
 Let us stress that there is no room for such connection unless higher symmetries are
 around. On the other hand, from this perspective (some of) HS symmetries acquire a
 nontrivial geometric meaning as relating nonequivalent cycles.
 An interesting remaining  question is whether it is possible for a given parameter $\eta$ for
 the charge $\langle\sigma_0,\eta\rangle$ to find $\eta^\prime$ such that
 $\langle\sigma_0,\eta\rangle = \langle\Sigma,\eta^\prime\rangle$ for a higher cycle $\Sigma$.
 Expression \eqref{eq:conjugate} gives only sufficient condition for this to be true.

In accordance with the Noether's theorem, quantized conserved charges resulting from the
lower fundamental cycle generate symmetry transformations
\eqref{eq:symmetries_Fock_r1_periodic}  of quantized rank-one fields
via the commutator. Charges are parametrized by elements of the star-product
algebra \eqref{eq:Moyal} which are conveniently packed into generating functions
\eqref{eq:sin_elements}
 closed under the star-product commutator as in \eqref{eq:sin_algebra}
and which subalgebra \eqref{eq:sin_algebra_even} resembles sine algebra introduced in \cite{Fairley}.
The Lie algebra \eqref{eq:sin_algebra} represents the full residual global symmetry of the  unfolded system
\eqref{eq:unfolded_r1} after imposing periodic conditions on the spinor variables $Y^A$.

The results of this paper may have several applications mentioned in Introduction.
One related to black hole solutions in the HS theory seems to be the most interesting.
We hope to consider this issue in the future.

\section*{Acknowledgments}
We are grateful to Olga Gelfond  for stimulating discussions.


\begin{thebibliography}{11}

\bibitem{Fronsdal:1985pd}
  C.~Fronsdal,
  ``Massless Particles, Orthosymplectic Symmetry And Another Type Of Kaluza-klein Theory,''
  In *Fronsdal, C. ( Ed.): Essays On Supersymmetry*, 163-265 and Calif. Univ. Los Angeles - UCLA-85-TEP-10 (85,REC.JUN.) 111 P. (508632)

\bibitem{Bandos:1999qf}
  I.~A.~Bandos, J.~Lukierski and D.~P.~Sorokin,
  ``Superparticle models with tensorial central charges,''
  Phys.\ Rev.\ D {\bf 61}, 045002 (2000)
  [hep-th/9904109v1].

\bibitem{Bandos:1999pq}
  I.~A.~Bandos, J.~Lukierski, C.~Preitschopf and D.~P.~Sorokin,
  ``OSp supergroup manifolds, superparticles and supertwistors,''
  Phys.\ Rev.\ D {\bf 61}, 065009 (2000)
  [hep-th/9907113v1].

\bibitem{Vasiliev:2001zy}
  M.~A.~Vasiliev,
  ``Conformal higher spin symmetries of 4-d massless supermultiplets and osp(L,2M) invariant equations in generalized (super)space,''
  Phys.\ Rev.\ D {\bf 66}, 066006 (2002)
  [hep-th/0106149].

\bibitem{Vasiliev:2001dc}
  M.~A.~Vasiliev,
  ``Relativity, causality, locality, quantization and duality in the Sp(2M) invariant generalized space-time,''
  In *Olshanetsky, M. (ed.) et al.: Multiple facets of quantization and supersymmetry* 826-872
  [hep-th/0111119].

\bibitem{Vasiliev:2002fs}
  M.~A.~Vasiliev,
  ``Higher spin conserved currents in Sp(2M) symmetric space-time,''
  Russ.\ Phys.\ J.\  {\bf 45}, 670 (2002)
  [Izv.\ Vuz.\ Fiz.\  {\bf 2002N7}, 23 (2002)]
  [hep-th/0204167].

\bibitem{Bandos:2002te}
  I.~A.~Bandos,
  ``BPS preons and tensionless superp-brane in generalized superspace,''
  Phys.\ Lett.\ B {\bf 558}, 197 (2003)
  [hep-th/0208110].

\bibitem{Didenko:2003aa}
  V.~E.~Didenko and M.~A.~Vasiliev,
  ``Free field dynamics in the generalized AdS (super)space,''
  J.\ Math.\ Phys.\  {\bf 45}, 197 (2004)
  [hep-th/0301054].

\bibitem{PluSorTsu}
  M.~Plyushchay, D.~Sorokin and M.~Tsulaia,
  ``Higher spins from tensorial charges and OSp(N|2n) symmetry,''
  JHEP {\bf 0304}, 013 (2003)
  [hep-th/0301067].

\bibitem{spspace}
  M.~A.~Vasiliev,
  ``Higher spin theories and Sp(2M) invariant space-time,''
  hep-th/0301235.

\bibitem{BanPasSorTon}
  I.~Bandos, P.~Pasti, D.~Sorokin and M.~Tonin,
  ``Superfield theories in tensorial superspaces and the dynamics of higher spin fields,''
  JHEP {\bf 0411}, 023 (2004)
  [hep-th/0407180].

\bibitem{BanBekAzSorTsu}
  I.~Bandos, X.~Bekaert, J.~A.~de Azcarraga, D.~Sorokin and M.~Tsulaia,
  ``Dynamics of higher spin fields and tensorial space,''
  JHEP {\bf 0505}, 031 (2005)
  [hep-th/0501113].

\bibitem{IvLuk}
  E.~Ivanov and J.~Lukierski,
  ``Higher spins from nonlinear realizations of OSp(1|8),''
  Phys.\ Lett.\ B {\bf 624}, 304 (2005)
  [hep-th/0505216].

\bibitem{Iv}
  E.~Ivanov,
  ``Nonlinear Realizations in Tensorial Superspaces and Higher Spins,''
  hep-th/0703056 [hep-th].

\bibitem{theta}
  O.~A.~Gelfond and M.~A.~Vasiliev,
  ``Higher Spin Fields in Siegel Space, Currents and Theta Functions,''
  JHEP {\bf 0903}, 125 (2009)
  [arXiv:0801.2191 [hep-th]].

\bibitem{sph}
  O.~A.~Gelfond and M.~A.~Vasiliev,
  ``Sp(8) invariant higher spin theory, twistors and geometric BRST formulation of unfolded field equations,''
  JHEP {\bf 0912}, 021 (2009)
  [arXiv:0901.2176 [hep-th]].

\bibitem{twistors}
  O.~A.~Gelfond and M.~A.~Vasiliev,
  ``Operator algebra of free conformal currents via twistors,''
  Nucl.\ Phys.\ B {\bf 876}, 871 (2013)
  [arXiv:1301.3123 [hep-th]].

\bibitem{Sorokin}
  I.~Florakis, D.~Sorokin and M.~Tsulaia,
  ``Higher Spins in Hyperspace,''
  JHEP {\bf 1407}, 105 (2014)
  [arXiv:1401.1645 [hep-th]].

\bibitem{rank}
  O.~A.~Gelfond and M.~A.~Vasiliev,
  ``Higher rank conformal fields in the Sp(2M) symmetric generalized space-time,''
  Theor.\ Math.\ Phys.\  {\bf 145}, 1400 (2005)
  [Teor.\ Mat.\ Fiz.\  {\bf 145}, 35 (2005)]
  [hep-th/0304020].

\bibitem{GSV}
  O.~A.~Gelfond, E.~D.~Skvortsov and M.~A.~Vasiliev,
  ``Higher spin conformal currents in Minkowski space,''
  Theor.\ Math.\ Phys.\  {\bf 154}, 294 (2008)
  [hep-th/0601106].

\bibitem{ads}
  O.~A.~Gelfond and M.~A.~Vasiliev,
  ``Conserved higher-spin charges in $AdS_4$,''
  Phys.\ Lett.\ B {\bf 754}, 187 (2016)
  [arXiv:1412.7147 [hep-th]].

\bibitem{Fairley} Fairlie D.B., Fletcher P., Zachos C.K., ``Trigonometric structure constants for new infinite-dimensional algebras'', Phys.\ Lett.\ B {\bf 218}, 203 (1989).

\bibitem{NonLinHSmanual}
  X.~Bekaert, S.~Cnockaert, C.~Iazeolla and M.~A.~Vasiliev,
  ``Nonlinear higher spin theories in various dimensions,''
  [hep-th/0503128].

\bibitem{Mumford} D.~Mumford, ``Tata lectures on theta I,'' Birkhäuser Boston (1983).

\bibitem{KZ}
  S.~Kharchev and A.~Zabrodin,
  ``Theta vocabulary II. Multidimensional case,''
  J.\ Geom.\ Phys.\  {\bf 104}, 112 (2016)
  [arXiv:1510.02699 [math.CV]].

\bibitem{BTZ}
  M.~Banados, C.~Teitelboim and J.~Zanelli,
  ``The Black hole in three-dimensional space-time,''
  Phys.\ Rev.\ Lett.\  {\bf 69}, 1849 (1992)
  [hep-th/9204099].


\bibitem{KlebPol}
  I.~R.~Klebanov and A.~M.~Polyakov,
  ``AdS dual of the critical O(N) vector model,''
  Phys.\ Lett.\ B {\bf 550}, 213 (2002)
  [hep-th/0210114].

\bibitem{Giombi:2012ms}
  S.~Giombi and X.~Yin,
  ``The Higher Spin/Vector Model Duality,''
  J.\ Phys.\ A {\bf 46} (2013) 214003
  [arXiv:1208.4036 [hep-th]].

\end{thebibliography}
\end{document}